\documentclass[aps,preprint]{revtex4}
\begin{document}
\title{Competition between Superconductivity and Charge Density
Wave Ordering in the Lu$_5$Ir$_4$(Si$_{1-x}$Ge$_x$)$_{10}$ Alloy System}
\author{Yogesh Singh, R. Nirmala, S. Ramakrishnan and S. K. Malik}
\address{Tata Institute of Fundamental Research, Mumbai-400005, India}
\begin{abstract}
\noindent 
We have performed bulk measurements such as dc magnetic susceptibility, electrical resistivity and heat capacity on the pseudo-ternary alloys Lu$_5$Ir$_4$(Si$_{1-x}$Ge$_x$)$_{10}$ to study the interplay and competition between superconductivity and the charge density wave (CDW) ordering transition. We track the evolution of the superconducting transition temperature T$_{SC}$ and the CDW ordering temperature T$_{CDW}$ as a function of x (concentration of Ge) ($0.0 \leq x\leq~1.0$). We find that increasing x (increasing disorder) suppresses the T$_{CDW}$ rapidly with the concomitant increase in T$_{SC}$. We present a temperature-concentration (or volume) phase diagram for this system and compare our results with earlier work on substitution at the Lu or Ir site to show how dilution at the Si site presents a different situation from these other works. The heat capacity data in the vicinity of the CDW transition has been analyzed using a model of critical fluctuations in addition to a mean-field contribution and a smooth lattice background. We find that the critical exponents change appreciably with increasing disorder. This analysis suggests that the strong-coupling and non mean-field like CDW transition in the parent compound Lu$_5$Ir$_4$Si$_{10}$ changes to a mean-field like transition with increasing Ge concentration.

\vskip 1truecm 
\noindent     
Ms number ~~~~~~~~~~~~PACS number:~71.45.Lr, 75.40.-s, 71.20.Lp, 72.15.-v\\
\end{abstract}
\maketitle
\newpage
\section{Introduction}
\label{sec:INTRO}
\noindent
Superconductivity (SC) and Charge Density Wave (CDW) ordering are two very different cooperative phenomena both of which occur due to Fermi Surface (FS) instabilities and require a large density of states (DOS) at the FS. Both phenomena involve an opening up of a gap at the FS and hence lead to a reduction in the DOS at the FS below their respective transition temperatures. It is therfore of great interest to investigate the effect the SC and CDW have on each other when both happen to occur in the same system. Several such studies, both experimental \cite{r0, r00, r000} and theoretical \cite{r01, r02}, have been undertaken in the past. However, most of these have been on systems with quasi-low-dimensional structures. 
Recently the compounds of the intermetallic series R$_5$Ir$_4$Si$_{10}$ (R~=~rare earth) which have essentially a 3-dimensional structure provide an opportunity where either superconductivity and CDW ordering or magnetism and CDW ordering co-exist in the same material depending on whether the rare-earth element R is nonmagnetic or magnetic \cite{r1,r2,r3,r4,r5}.\\
The compounds Lu$_5$Ir$_4$Si$_{10}$ and Lu$_5$Ir$_4$Ge$_{10}$, like the other compounds of this series, crystallize in the tetragonal Sc$_5$Co$_4$Si$_{10}$ (space group Pm3n) type structure. Lu$_5$Ir$_4$Si$_{10}$ is known to superconduct below 3.9~K and it has also been shown to exhibit a strongly coupled charge density wave transition below T$_{CDW}$~=~83~K \cite{r1,r2,r3}. It has been shown that the CDW transition gets progressively suppressed to lower temperatures on the application of hydrostatic pressure and is completely suppressed at a critical pressure of 21~kbar. There is a sudden enhancement of the superconducting transition temperature T$_{SC}$ from 3.9~K to 9~K at this critical pressure \cite{r2}. This implies an intricate interplay and competition between the two phenomena. From heat capacity and susceptibility measurements, almost a 36\% reduction in the density of states at the Fermi level due to the CDW transition has been estimated \cite{r2}. This observation was further supported by a $^{175}$Lu NMR measurement where an abrupt dip in the Knight shift (which is in general proportional to the DOS) at 83~K indicated a loss of density of states at the Lu site \cite{r6}. However, similar $^{29}$Si NMR measurements performed recently on polycrystalline samples of Lu$_5$Ir$_4$Si$_{10}$ suggest that there is no loss of density of states at the Si site across the CDW transition \cite{r7}. Thus given that pressure suppresses the T$_{CDW}$ (since the elastic energy cost to distort the lattice is now enhanced and the gain in electronic energy due to the gapping at the FS can over-ride this cost at lower temperatures only \cite{r8}) and that it seems that the Si site is not involved in the CDW distortions, one would expect the isostructural Lu$_5$Ir$_4$Ge$_{10}$ compound which has a larger unit cell, to undergo a CDW ordering at an elevated temperature compared to the Si compound. However, we find that a high quality sample ($R(300K)\over R(4K)$~=~98) of Lu$_5$Ir$_4$Ge$_{10}$ only undergoes a transition into the superconducting state below 2.4~K without displaying any CDW transition at higher temperatures.\\ 
Hence, it is of interest to study the evolution of the superconductivity and the CDW transition when we substitute small quantities of Ge for Si in Lu$_5$Ir$_4$Si$_{10}$. Towards this end, we have carried out a detailed investigation of the superconductivity and CDW ordering in the alloy system Lu$_5$Ir$_4$(Si$_{1-x}$Ge$_x$)$_{10}$ via our dc magnetic susceptibility, electrical resistivity and heat capacity measurements.

\par

\section{EXPERIMENTAL DETAILS}
\label{sec:EXPT}
\noindent
Polycrystalline samples (10~gms each) of the parent compounds Lu$_5$Ir$_4$Si$_{10}$ and Lu$_5$Ir$_4$Ge$_{10}$ were prepared first by arc melting the constituent elements in stoichiometric proportions. The purity of the Lu was 99.99\%, that of Ir was 99.9\% and that of Si and Ge was 99.999\%. The melted ingots were flipped over and remelted 6 to 8 times to ensure homogenous mixing of the constituents. These ingots were used as master alloys for the preparation of the pseudo-ternary alloys.
Polycrystalline samples of Lu$_5$Ir$_4$(Si$_{1-x}$Ge$_x$)$_{10}$ with x~=~0.0, 0.005, 0.01, 0.02, 0.05, 0.1, 0.2, 0.4, 1.0 were then prepared by arc-melting together pieces taken in appropriate proportions from the previously prepared master alloys of the parent compounds Lu$_5$Ir$_4$Si$_{10}$ and Lu$_5$Ir$_4$Ge$_{10}$. The samples were annealed in a sealed quartz tube at 950~$^o$C for 10 days. Powder X-ray diffraction measurements confirmed the structure and the absence of any impurity phases. The lattice constants for all the samples were estimated from a least squares fit of their X-ray diffraction patterns. The results are given in Table 1. The lattice constants and the lattice volume increase roughly linearly as the Ge concentration is increased in the alloy.

The dc magnetic susceptibility in the temperature range 2~K to 300~K was measured using a commercial superconducting quantum interference device (SQUID) magnetometer (MPMS, Quantum design, USA). Electrical resistivity between 1.5~K and 300~K was measured with an LR-700 ac resistance bridge (Linear Research, USA) using the 4-probe technique with electrical contacts made by silver paste on bar shaped slides cut from the annealed samples. The heat capacity was measured between 2~K and 150~K using a commercial physical property measurement system (PPMS, Quantum Design, USA). The superconducting transition temperature and the CDW ordering temperature were determined by peaks in the derivatives of the magnetic susceptibility (d$\chi$/dT) and electrical resistivity (d$\rho$/dT) vs temperature data and by the corresponding peaks in the heat capacity data obtained after subtracting a smooth lattice background.

\section{RESULTS}
\label{sec:RES}
\subsection{Magnetic Susceptibility studies}
\noindent
In Fig.~1, we show the temperature dependence of the dc susceptibility for the samples Lu$_5$Ir$_4$(Si$_{1-x}$Ge$_x$)$_{10}$ with x~=~0.0, 0.005, 0.01, 0.02, 0.05 and 0.1 between 1.7~K and 300~K to concentrate on the samples with those values of x for which the CDW ordering transition is seen. The left hand panels in the figure show the zero field cooled (ZFC) and field cooled (FC) data recorded in a field of 10~Oe in the low temperature range to observe the superconducting transition. The right hand panels show the susceptibility data between 5~K and 300~K recorded in a field of 1~koe to look for the CDW ordering transition.
The signature of a CDW in the susceptibility of a nonmagnetic sample is a diamagnetic drop across the transition as the sample is cooled into the CDW state. This comes about due to the reduction in the density of states at the Fermi surface because of the opening up of a gap at the Fermi surface accompanying the CDW ordering.
It can immediately be seen that even small Ge concentrations affect the CDW strongly. From an onset temperature of 83~K for the unsubstituted sample Lu$_5$Ir$_4$Si$_{10}$, the CDW starts to shift to lower temperatures and also begins to broaden out considerably as the Ge concentration in the alloy increases. At a concentration of only 10\% of Ge, the CDW has been suppressed so much that it can no longer be detected. This can be seen in the lowest panel of Fig.~1. The values of the T$_{CDW}$ have been determined by peaks in the d($\chi$)/dT vs T plots. The samples with higher concentration of Ge (x$>$0.02) also do not show any signature of the CDW transition as can be seen from Fig.~2.\\
Importantly, there is a simultaneous increase in the superconducting transition temperature T$_C$ as can be clearly seen in the left hand panels of Figs.~1 and 2. The superconducting transition temperature reaches a maximum value of 6.6~K for x=0.2, {\sl i.e.}, for the sample Lu$_5$Ir$_4$Si$_8$Ge$_2$ (see Fig.~2). It must be noted that the CDW is last seen for x=0.05 but the superconducting transition temperature increases til the x=0.20 sample (see Fig.~2). We will return to this point when we discuss our results. For samples with higher concentrations of Ge, the T$_C$ reduces until it reaches a value of 2.4~K for the pure Ge sample Lu$_5$Ir$_4$Ge$_{10}$.\\ 
The upturn in the susceptibility of all samples at the lowest temperatures is probably due to trace amounts (few ppm) of paramagnetic impurities in the samples. However, as we increase the concentration of Ge, we find that there is a tail in the susceptibility for temperatures higher than T$_{CDW}$ (see fig.~1 and 2). This tail is not seen for the pure samples Lu$_5$Ir$_4$Si$_{10}$ or Lu$_5$Ir$_4$Ge$_{10}$. This indicates that the high temperature tails in the Ge substituted samples are not due to the increasing presence of any paramagnetic impurities introduced due to Ge substitution. Therefore, we believe that these tails in the Ge substituted samples are arising due to the presence of regions with a distribution of T$_{CDW}$'s. These regions however, have to be small since prominent anomalies occur only at lower temperatures where the bulk of the sample undergoes the CDW transition.

\subsection{Resistivity Studies}
\noindent
In Fig.~3, we show the temperature dependence of the normalized resistance for the Lu$_5$Ir$_4$(Si$_{1-x}$Ge$_x$)$_{10}$ samples with x~=~0.0, 0.005, 0.01, 0.02 and 0.05. The main panels show the data between 1.6~K and 300~K displaying the CDW transition at higher temperatures seen as an abrupt increase in the resistance across the transition. The insets show the low temperature data between 1.6~K and 10~K on an expanded scale to highlight the superconducting transition for the various samples. For the unsubstituted sample Lu$_5$Ir$_4$Si$_{10}$, the CDW transition is seen as an abrupt step like increase in the resistance (see top right panel of Fig.~3). This upturn occurs due to the opening up of a gap at the Fermi surface (FS) associated with the CDW transition. However, the resistance behavior after the transition remains metallic indicating that only partial gapping of the FS occurs.\\ 
The CDW, which is seen as a sharp and abrupt step like increase in the resistance for Lu$_5$Ir$_4$Si$_{10}$, is suppressed to lower temperatures on substituting with Ge. Only a 0.5 atomic percent of Ge reduces the CDW transition temperature T$_{CDW}$ to 75~K from 83~K for the unsubstituted sample. The CDW transition is also progressively broadened or smeared out and the magnitude of the anomaly in the resistance across the transition also reduces considerably. All these facts indicate a strong weakening/suppression of the CDW transition as the amount of Ge in the alloy Lu$_5$Ir$_4$(Si$_{1-x}$Ge$_x$)$_{10}$ is increased. \\
To get a quantitative estimate of the effect of increasing disorder on the CDW transition, we list in Table~2 the values of the CDW transition temperature determined by the peak position in the plots of d$\rho$(T)/dT vs T (not shown here) for various values of x (concentration of Ge in the alloy). 
Also listed in column 3 is the percentage increase in the resistance across the CDW transition as a measure of the magnitude of the resistive anomaly due to the CDW. Lastly in column 4, we have given the width of the transition which has been estimated from the full width at half maximum (FWHM) of the peaks at the transition as observed in the derivative of the resistivity data. No signatures of the CDW transition are observed for samples with higher Ge concentration (x$>$0.1) although the superconducting transition temperature keeps increasing upti the x=0.20 sample, as is shown in Fig.~4 for samples with x~=~0.1, 0.2, 0.4 and 1.0. The resistivity behavior of these samples resembles that of normal intermetallic alloys and it is interesting to note that the residual resistivity ratio (RRR~=~$R(300K)\over R(4K)$) increases for samples with larger Ge substitution as compared to the low concentration samples where there is an upturn in the resistivity due to the CDW. The superconducting transition temperature also ceases to increase and starts to reduce once the effect of the CDW is completely suppressed until it reaches a value of about 2.4~K for the end member Lu$_5$Ir$_4$Ge$_{10}$.\\ 
These data show that the resistive anomaly due to the CDW transition is strongly suppressed, smeared out or broadened and weakened by atomic disorder.\\
\subsection{Heat Capacity Studies}
\noindent
Fig.~5 shows the temperature dependence of the heat capacity for the compounds Lu$_5$Ir$_4$(Si$_{1-x}$Ge$_x$)$_{10}$ for x=~0.0, 0.005, 0.01, 0.02 and 0.05 between 2~K and 300~K. The insets show the low temperature behavior on an expanded scale to show the superconducting transitions for the various compounds. The top panel shows the heat capacity for the pure Si sample Lu$_5$Ir$_4$Si$_{10}$. The large peak of almost 75~J/mol~K at 81~K (over the considerable lattice heat capacity at these temperatures) denotes the transition of the compound into the CDW state. Large anomalies are also seen in the heat capacity for compounds with x~=~.005 and 0.01. However, they appear to be reduced in magnitude and broadened compared to the anomaly seen in the heat capacity for Lu$_5$Ir$_4$Si$_{10}$. For the compounds with x~=~0.02 and 0.05, we observe weak anomalies only after subtraction of the heat capacity of the compound Lu$_5$Ir$_4$Si$_6$Ge$_4$ (which does not undergo any CDW transition) from the total heat capacity. The heat capacity jumps C$_{CDW}$ at the CDW transition for all the compounds have been obtained in a similar way and the entropy change S$_{CDW}$ across the CDW transition has been estimated by integrating the C$_{CDW}$/T versus T curves. This is shown in Fig.~6 for the samples which undergo the CDW transition. The subtraction of the lattice heat capacity is not perfect and this shows up in the small tails observed below T$_{CDW}$ in the C$_{CDW}$ plots for the samples with x~=~0.02 and 0.05 (fig.~6). Hence, the values of the entropy associated with the CDW transitions is slightly overestimated for the x~=~0.02 and 0.05 samples. Nevertheless, it can be seen that the entropy change associated with the CDW transitions is considerable. We have collected the specific heat parameters for the CDW transition for each compound in Table~4. \\
The CDW transition in Lu$_5$Ir$_4$Si$_{10}$ has been shown to be of strong coupling and non mean-field like nature \cite{r3,r9}. In particular, Kuo {\it et al} \cite{r9} have analyzed the specific heat of Lu$_5$Ir$_4$Si$_{10}$ in the vicinity of the CDW transition with a model of critical fluctuations in addition to the BCS mean-field contributions. They find critical exponents which are much larger ($\sim$2) than expected for a mean-field like transition (0.5). They also estimate the effective electronic specific heat coefficient $\gamma^*$ and compare it with the bare Sommerfeld's coefficient $\gamma$ to show that the transition is of a strong-coupling nature \cite{r9}. We have also analyzed our specific heat data in a similar manner for the compounds which show the CDW transition to see how these parameters evolve as the CDW is suppressed by the increasing disorder. We would like to understand the effect of disorder on the first-order like CDW transition in the parent compound Lu$_5$Ir$_4$Si$_{10}$ and we hope to get insight about this in the way the critical exponents change for the samples with increasing Ge concentration. Similarly $\gamma^*$T$_{CDW}$, which is an estimate of the specific heat jumps at T$_{CDW}$ is expected to decrease.
Therefore, using the notation of ref. 8, the specific heat can be written as
$$C~=~C_L~+~C_{MF}~+~C_{fl} \eqno {(1)}$$
where C$_L$ is the lattice background, C$_{MF}$ is the mean-field term below T$_{CDW}$ and C$_{fl}$ is the contribution to the heat capacity from critical fluctuations. At these temperatures the lattice term can be assumed to take the from given by the Einstein's model 
$$C_L~=~a_1({a_2\over T})^{a_3}{e^{a_1/T}\over {(e^{a_2/T}-1)}^2} \eqno{(2)}$$
Thus, the total heat capacity above and below T$_{CDW}$ was fitted to the functional form 
$$C^-~=~C_L~+~\gamma^*T_{CDW}(1+\beta t)~+~b^-|t|^{-\alpha^-},  ~~~~~T~<~T_{CDW}$$
$$C^+~=~C_L~+~b^+|t|^{-\alpha^+},  ~~~~~T~>~T_{CDW} \eqno{(3)}$$
Here, the mean-field term below T$_{CDW}$ is given by
$$C_{MF}~=~\gamma^*T_{CDW}(1+\beta t) \eqno{(4)}$$
where $\gamma^*$ is the effective electronic specific heat coefficient.\\
The critical fluctuation contribution to the heat capacity has been given by
$$C_{fl}^-~=~b^-|t|^{-\alpha^-},  ~~~~T~<~T_{CDW} $$
$$C_{fl}^+~=~b^+|t|^{-\alpha^+},  ~~~~T>~T_{CDW} \eqno{(5)}$$
Here, $a_1$, $a_2$, $a_3$, $\gamma^*$, $\beta$, $b^-$ and $b^+$ are the effective fitting parameters, $\alpha^-$ and $\alpha^+$ are called the critical exponents and $t~=~({T_{CDW}-T\over T_{CDW}})$ is the reduced temperature.
Following the procedure of ref. 8, we have fitted the heat capacity data for the samples Lu$_5$Ir$_4$(Si$_{1-x}$Ge$_x$)$_{10}$ with x~=~0.0, 0.005, 0.01, 0.02 and 0.05. A very good agreement between the fit and the actual data is obtained. This is shown in Fig.~7 for Lu$_5$Ir$_4$Si$_{10}$ and Lu$_5$Ir$_4$Si$_{9.8}$Ge$_{0.2}$. The lower panel in this figure shows that for the sample Lu$_5$Ir$_4$Si$_{9.8}$Ge$_{0.2}$, although a finite discontinuity is observed at T$_{CDW}$ in the fit, the data only shows a disorder/inhomogenity broadened weak anomaly. Therefore, it can be argued that although the estimates of the critical exponents for the samples with x~=0.02 and 0.05 may have larger error bars, they may not be completely incorrect given that the evolution of the exponents for these two compounds follow the same trend which had emerged for the lower Ge concentration sample.\\ 
The fitted parameters for all the samples are collected in Table~4.
As expected, the parameters of the lattice background term do not vary too much. It is seen that the critical exponents $\alpha^-$ and $\alpha^+$ for Lu$_5$Ir$_4$Si$_{10}$ are close to 2 (in agreement with the previous report \cite{r9}) and are much larger than expected (=~0.5) for a mean-field like transition. However, with increasing Ge concentration in the alloy, the values of the critical exponents reduce progressively until they reach a value of about 0.6 for the compound Lu$_5$Ir$_4$Si$_{9.5}$Ge$_{0.5}$. The value of the electronic specific heat $\gamma^*$ comes out to be $7.2\times10^{-2}(J/mol K^2)$ for Lu$_5$Ir$_4$Si$_{10}$. The bare Sommerfeld's coefficient $\gamma$ for Lu$_5$Ir$_4$Si$_{10}$ is $9.2\times10^{-3}(J/mol K^2)$ \cite{r9} which gives $\gamma^*\over \gamma$~=~7.85 which is about 5.5 times larger than the BCS weak-coupling limit value of 1.43. This suggests the strong coupling nature of the CDW transition. The value of $\gamma^*$ and $\gamma^*T_{CDW}$ (which is an indication of the specific heat jump at the CDW transition) progressively decrease with increasing x (see mean-field column in table~4).
The evolution of the critical exponents and $\gamma^*$ suggests that the CDW transition is of a strongly coupled nature and non mean-field like in the pure Si compound Lu$_5$Ir$_4$Si$_{10}$ but changes to a mean-field like transition with increasing disorder.   

\section{Discussion: Interplay Between Superconductivity and Charge density Wave Ordering}
\label{sec:DIS}
\noindent
In Fig.~8, we combine the data from our magnetic susceptibility and electrical resistivity measurements into a temperature-concentration (T-x) phase diagram for the alloys Lu$_5$Ir$_4$(Si$_{1-x}$Ge$_x$)$_{10}$. The top panel shows the superconducting transition temperature vs. concentration for all samples. The bottom panel shows T$_C$ and T$_{CDW}$ vs. concentration plots for samples with x=~0.005, 0.01, 0.02, 0.05 and 0.1 to highlight the relation between T$_C$ and T$_{CDW}$. It can be seen that for low impurity concentration (x$\leq$~0.02) the suppression of the CDW transition temperature T$_{CDW}$ and the enhancement of the superconducting transition temperature T$_C$ is quasi-linear. We determine the initial concentration dependence (for small x) of T$_{CDW}$ and T$_C$ to be dT$_{CDW}$/dx~=~-16.2($\pm$0.5)~K/at.\% and dT$_C$/dx~=~1.0($\pm$0.1)~K/at.\%. Interestingly, our values differ from the concentration dependences of these transition temperatures when the substitution is done at the Lu or Ir site. For instance, for the pseudo-ternary compound (Lu$_{1-x}$Sc$_x$)$_5$Ir$_4$Si$_{10}$, it was found that dT$_{CDW}$/dx~=~-18.5~K/at.\% and dT$_C$/dx~=~0.5~K/at.\% \cite{r10}. Thus, the initial suppression of T$_{CDW}$ for our compounds is 2~K slower and the enhancement of T$_C$ is double that of the (Lu$_{1-x}$Sc$_x$)$_5$Ir$_4$Si$_{10}$ alloys. Also, for the pseudo-ternary alloys obtained by substituting Sc at the Lu or Rh at the Ir site, no CDW is observed at 5\% substitution while we still observe a CDW transition as high as 47~K for the 5\% sample. Only at a substitution of 10\% do we fail to observe any signature of a CDW transition. However, it would be wrong to deduce that the CDW, which occured as high as 47~K (albeit very weak) for x~=~5\%, suddenly disappears at x~=~10\%. It must be noted that the superconductivity is enhanced to 6.3~K for the 10\% Ge sample and in fact attains a maximum value equal to 6.6~K for a 20\% Ge substitution (see top panel of Fig.~5). This indicates that altough we are unable to detect any signatures of the CDW (probably because it had become extremely weak) in our bulk measurements on samples with Ge concentrations higher than 5\%, some traces of the CDW ordering persist even for samples with higher concentration of Ge and the gradual suppression of this by increasing disorder enhances the superconductivity to as high as 6.6~K for the 20\% Ge sample. These values are about 0.5~K to 1~K higher than the superconductivity enhancement observed at the suppression of the CDW transition in pseudo-ternary compounds obtained with substitution at the Lu or Ir site \cite{r10}. All these results indicate that introducing Ge at the Si site is indeed different from substituting at the Lu or Ir sites.

Let us now look at the concentration-temperature phase diagram of Lu$_5$Ir$_4$(Si$_{1-x}$Ge$_x$)$_{10}$ to understand the interplay between the superconductivity and charge density wave ordering in this system. The lower panel clearly shows that the CDW ordering is strongly suppressed by even small impurity concentrations until there is no signature of a CDW transition in either the magnetic or the transport measurements for the samples with 10\% Ge concentration. Importantly, the superconductivity is simultaneously enhanced. The superconducting transition temperature T$_C$ increases rapidly for small Ge concentrations where the CDW is also affected the most. At the value of x~=~0.1, for which the CDW has been completely suppressed, the T$_C$ ceases to increase rapidly and goes through a broad maximum at a value of about 6.6~K before decreasing again for higher values of x where the disorder takes over. This strongly indicates that the superconductivity in this system is enhanced at the expense of the CDW ordering. Also, we note that the CDW is suppressed even though we are expanding the lattice which suggests that disorder suppresses the CDW more strongly than pressure.
  
\section{CONCLUSION}
\label{sec:CON}
We have measured dc magnetic susceptibility, electrical resistivity and heat capacity of the allow system Lu$_5$Ir$_4$(Si$_{1-x}$Ge$_x$)$_{10}$ for x~=~0.0, 0.005, 0.01, 0.02, 0.05, 0.1, 0.2, 0.4, and 1.0 to investigate the evolution of the superconductivity and CDW transitions with increasing disorder. We find that the CDW transition is strongly suppressed from 83~K for x~=~0 down to 47~K for x = 0.05. There is no signature of the CDW for higher values of x. There is a concomitant enhancement of the superconducting transition temperature from 3.9~K for x~=~0.0 to 6.6~K for x~=~0.2. Therefore, our results indicate that there is a strong interplay and competition between the superconductivity and the CDW ordering in this system. We also conclude that the CDW transition is more sensitive to disorder than to pressure since the CDW is suppressed even though we are expanding the lattice (which should normally have taken the CDW to higher temperatures).\\
Analysis of the heat capacity data in the vicinity of the CDW phase transition suggests that disorder drives the strongly-coupled and non mean-field like CDW transition in Lu$_5$Ir$_4$Si$_{10}$ to a broadened mean-field like transition in the samples where Ge is substituted for Si.

\begin{table}
\centering
\caption{{ Lattice parameters of Lu$_5$Ir$_4$(Si$_{1-x}$Ge$_x$)$_{10}$.}}
\begin{tabular}{|c|c|c|c|c|c|}
\hline 
x&a=b($\pm$.0005)~($\AA$)&c($\pm$.0005)~($\AA$)&v~($\AA$)& c/a \\
\hline 
\hline
0.00&12.4756&4.1767&650.0682&0.33482\\ 
0.005&12.4767&4.1770&650.2361&0.33479\\ 
0.01&12.4795&4.1775&650.5915&0.33475\\ 
0.02&12.4837&4.1781&651.1333&0.33469\\ 
0.05&12.4923&4.1804&652.381&0.33464\\ 
0.10&12.5103&4.1825&654.5845&0.33432\\ 
0.20&12.5427&4.1877&658.8020&0.33387\\ 
0.40&12.6082&4.2008&667.7873&0.33318\\ 
1.00&12.8281&4.2415&697.98187&0.33064\\ \hline
\end{tabular}
\label{table 1}
\end{table}

\begin{table}
\centering
\caption{{ CDW transition parameters obtained from the temperature dependence of the resistivity of Lu$_5$Ir$_4$(Si$_{1-x}$Ge$_x$)$_{10}$.}}
\begin{tabular}{|c|c|c|c|c|c|}
\hline 
x&T$_{CDW}$~(K)&$\Delta$$\rho$(T$_{CDW}$)/$\rho$(300~K)&$\Delta$T~(K)\\
\hline
\hline
0.00&83&32\%&2\\ 
0.005&75&19\%&3.2\\ 
0.01&64&8.5\%&7\\ 
0.02&51&3.2\%&13\\ 
0.05&47&2.6\%&24\\ 
0.10&No CDW&-&-\\ \hline
\end{tabular}
\label{table 2}
\end{table}

\begin{table}
\centering
\caption{{ Parameters obtained from the specific heat data of Lu$_5$Ir$_4$(Si$_{1-x}$Ge$_x$)$_{10}$.}}
\begin{tabular}{|c|c|c|c|c|c|}
\hline 
x&T$_{CDW}$~(K)&C$_{CDW}$~(J/mol~K)&S$_{CDW}$~(R)\\
\hline
\hline
0.00&81&8.5&0.42\\ 
0.005&75&6.8&0.46\\ 
0.01&64&3.9&0.7\\ 
0.02&52&1.5&0.44\\ 
0.05&47&0.75&0.35\\ \hline
\end{tabular}
\label{table 1}
\end{table}

\begin{table}
\centering
\caption{{ Parameters obtained from fitting the specific heat data of Lu$_5$Ir$_4$(Si$_{1-x}$Ge$_x$)$_{10}$ to a model of critical fluctuations plus a mean-field contribution.}}
\begin{tabular}{|c|ccc|cc|cccc|}
\hline 
&&Lattice term&&~~~~~~~mean-field term&&&&fluctuation term&\\
x&a$_1$(J/mol K)&a$_2$(K)&a$_3$&$\gamma^*$(J/mol K$^2$)&$\beta$&~b$^-$(J/mol K)&$\alpha^-$&~~b$^+$(J/mol K)&$\alpha^+$\\
\hline
\hline
0.00&346&261&1.37&7.21e-2&6.981&8.86e-3&1.77&8.8e-4&1.81\\ 
0.005&313&225&1.20&1.44e-2&7.65&1.49e-2&1.48&1.1e-2&1.42\\ 
0.01&310&225&1.22&6.25e-3&6.5&3.9e-3&1.14&1.69e-2&1.11\\ 
0.02&316&229&1.24&3.59e-3&3.7&-7.4e-4&0.92&1.1261e-4&0.92\\ 
0.05&326&236&1.25&9.04e-4&2.89&-1.8&0.72&2.73&0.49\\ \hline
\end{tabular}
\label{table 1}
\end{table}

\begin{figure}
\caption{The temperature dependence of the magnetic susceptibility for Lu$_5$Ir$_4$(Si$_{1-x}$Ge$_x$)$_{10}$ for x~=~0.0, 0.005, 0.01, 0.02, 0.05 and 0.1. The left hand panels show the low temprature behavior for all the samples to high light the superconducting transition. The right hand panels show the CDW transition. (see text for details).
\label{fsus1}}
\end{figure}
\begin{figure}
\caption{The temperature dependence of the magnetic susceptibility for Lu$_5$Ir$_4$(Si$_{1-x}$Ge$_x$)$_{10}$ for x~=~0.2, 0.4 and 1.0. The left hand panels show the low temprature behavior for all the samples to high light the superconducting transition. The right hand panels show the absence of CDW anomalies for these compounds.
\label{fsus2}}
\end{figure}
\begin{figure}
\caption{The temperature dependence of the electrical resistivity for Lu$_5$Ir$_4$(Si$_{1-x}$Ge$_x$)$_{10}$ for x~=~0.0, 0.005, 0.01, 0.02 and 0.05. The main panels show the data between 2~K and 300~K to highlight CDW transitions. The insets show the low temperature behavior to highlight the superconducting transition for all samples. The superconducting transition temoerature for each compound is given in the inset.
\label{fres1}}
\end{figure}
\begin{figure}
\caption{The temperature dependence of the electrical resistivity for Lu$_5$Ir$_4$(Si$_{1-x}$Ge$_x$)$_{10}$ for x~=~0.1, 0.2, 0.4 and 1.0. The insets show the low temperature behavior to highlight the superconducting transition for all samples. The superconducting transition temoerature for each compound is given in the inset.
\label{fres2}}
\end{figure}
\begin{figure}
\caption{The temperature dependence of the specific heat for Lu$_5$Ir$_4$(Si$_{1-x}$Ge$_x$)$_{10}$ for x~=~0.0, 0.005, 0.01, 0.02 and 0.05. The main panels show the data between 2~K and 300~K. The large peaks are signatures of the CDW ordering transitions. The insets show the low temperature behavior to highlight the BCS like superconducting transition for all samples.
\label{fhc1}}
\end{figure}
\begin{figure}
\caption{The temperature dependence of the excess specific heats for Lu$_5$Ir$_4$(Si$_{1-x}$Ge$_x$)$_{10}$ for x~=~0.0, 0.005, 0.01, 0.02 and 0.05 across the CDW ordering transiitons and the entropy involved in the transitions (see text for details). 
\label{fhc2}}
\end{figure}
\begin{figure}
\caption{The specific heat data (open circles) and the fit (solid lines) to the model of critical fluctuations plus a mean-field contribution. The top panel shows the fitting results for the sample Lu$_5$Ir$_4$Si$_{10}$ and the bottem panel shows the fitting results for the sample Lu$_5$Ir$_4$Si$_{9.8}$Ge$_{0.2}$ plotted as C/T vs T to bring out the weak anomaly at T$_{CDW}$ for this sample. (See text for details of the fitting).   
\label{fhc3}}
\end{figure}
\begin{figure}
\caption{The temperature-concentration phase diagram for the alloy system Lu$_5$Ir$_4$(Si$_{1-x}$Ge$_x$)$_{10}$ . The top panel shows the variation of superconducting transition temperature T$_{C}$ with concentration x for all the samples and the bottom panel shows the variatioon of T$_{C}$ and CDW ordering transition temperature T$_{CDW}$ with concentration x for the samples which show the CDW transition. 
\label{fcdw}}
\end{figure}
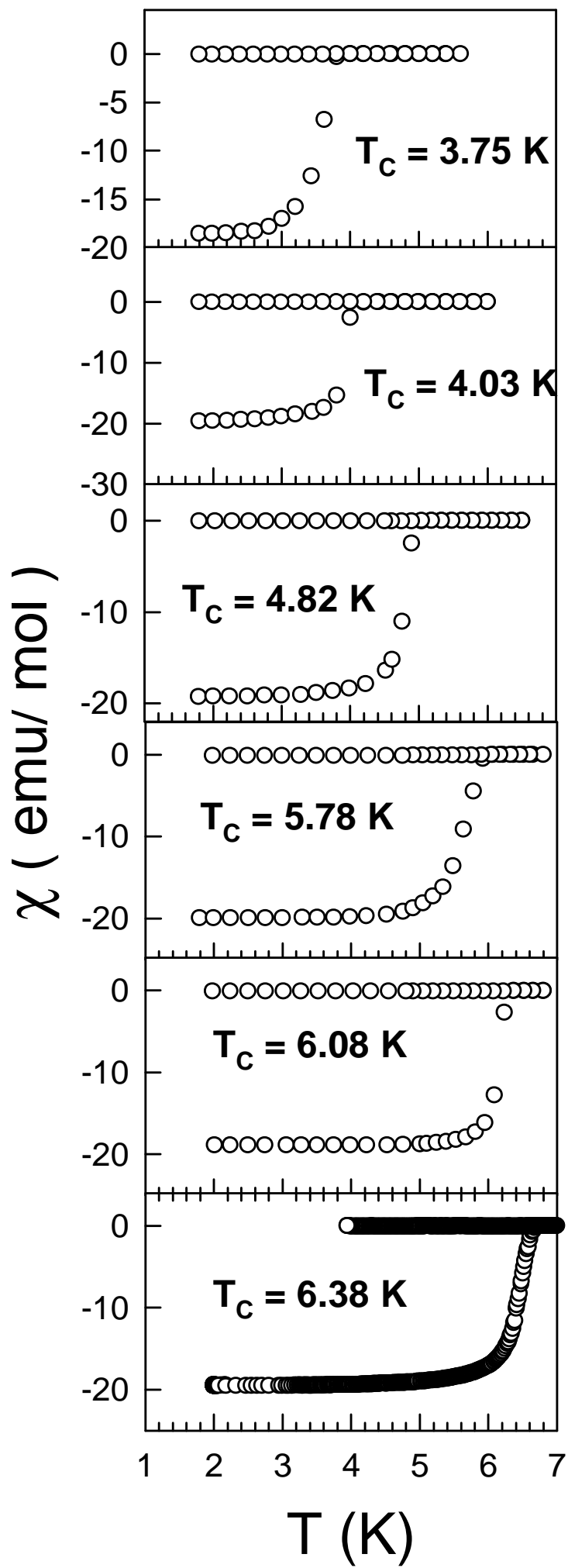
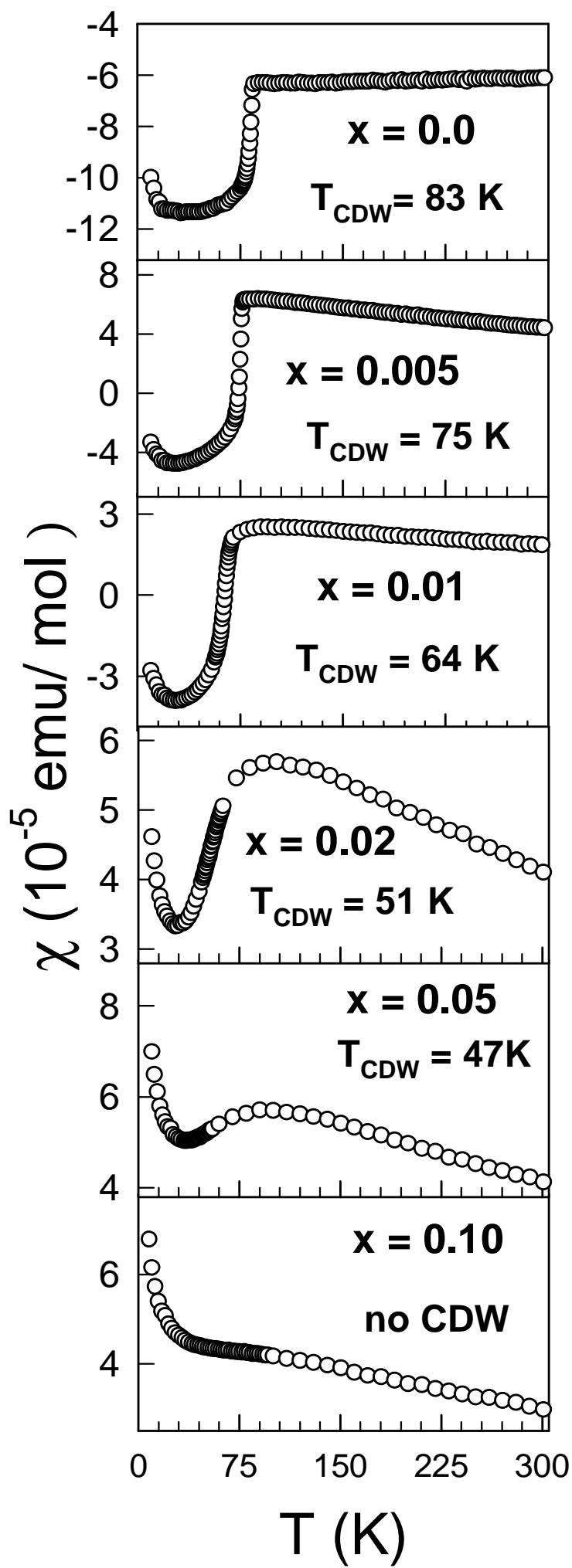
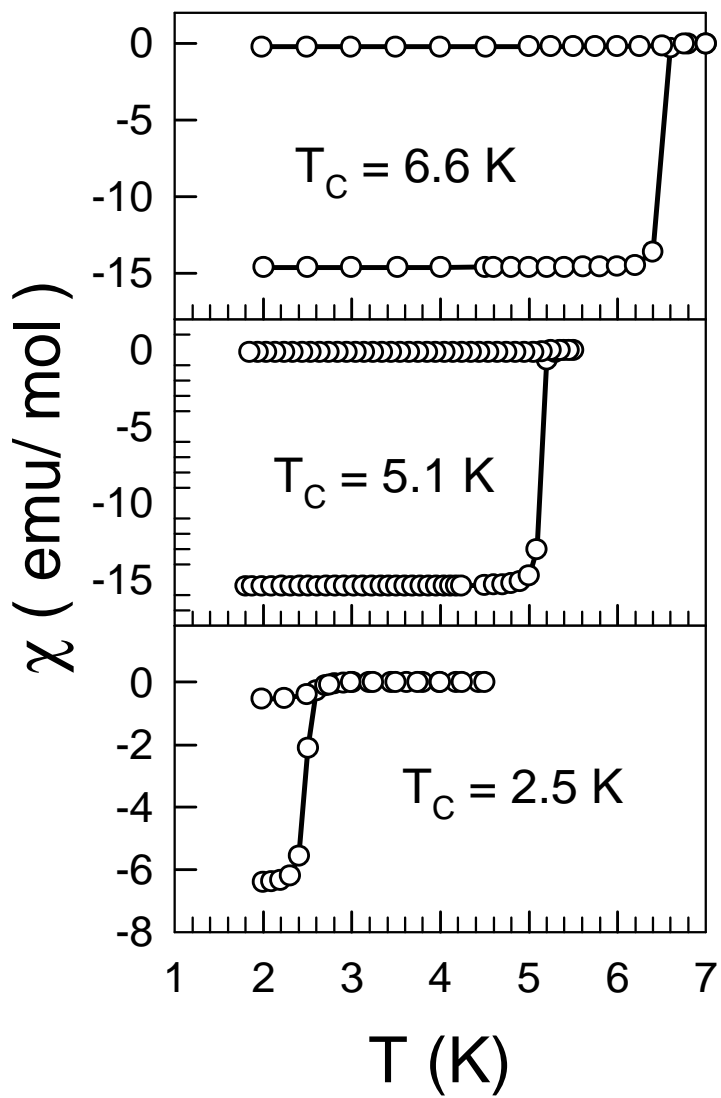
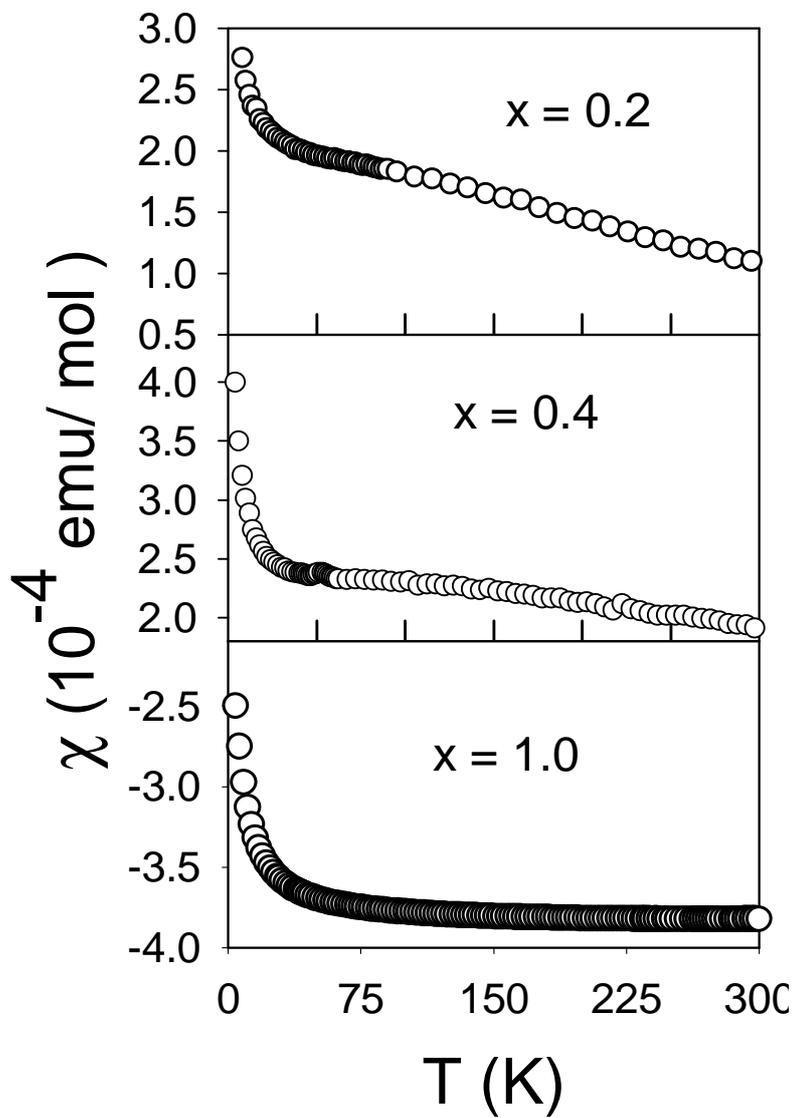
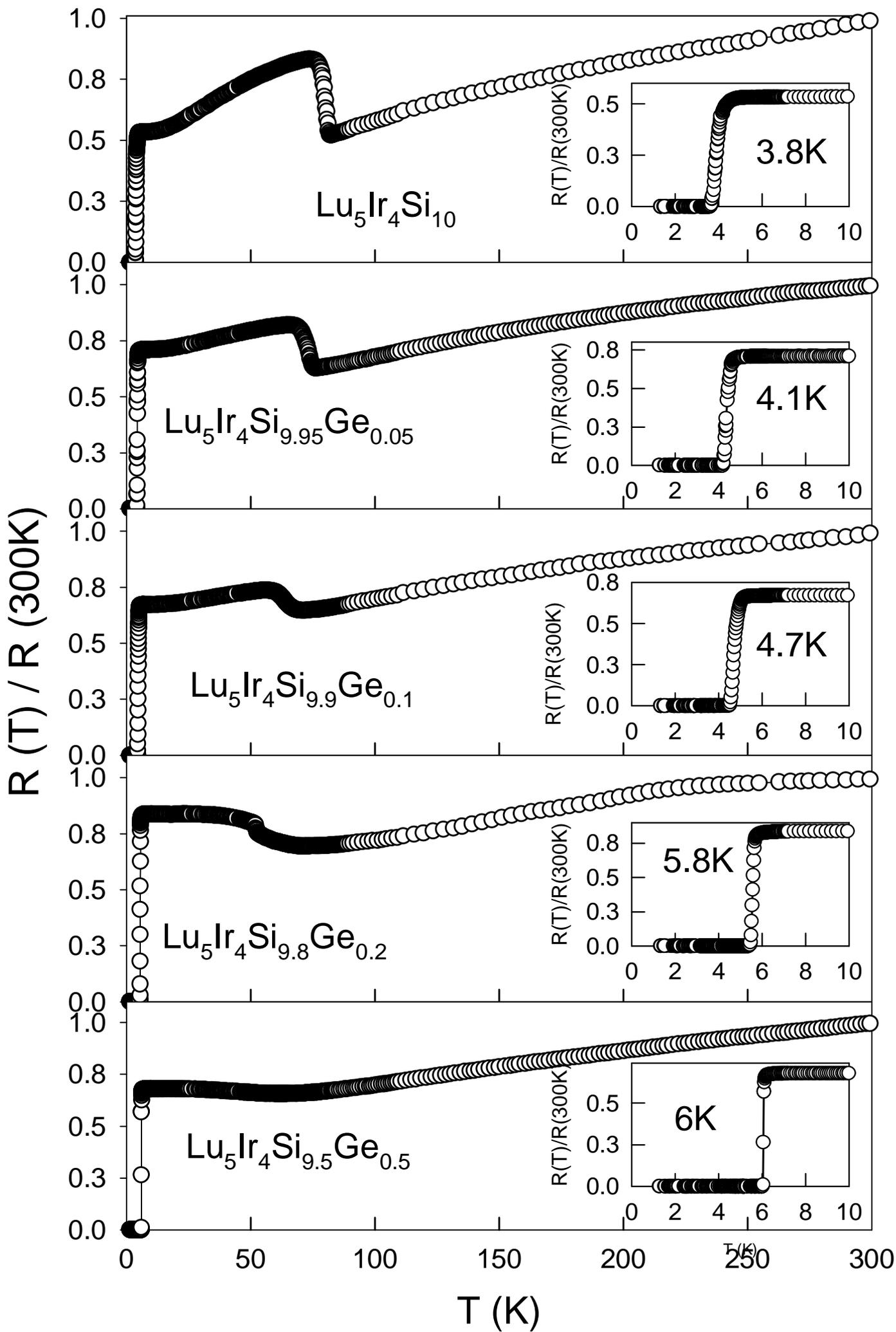
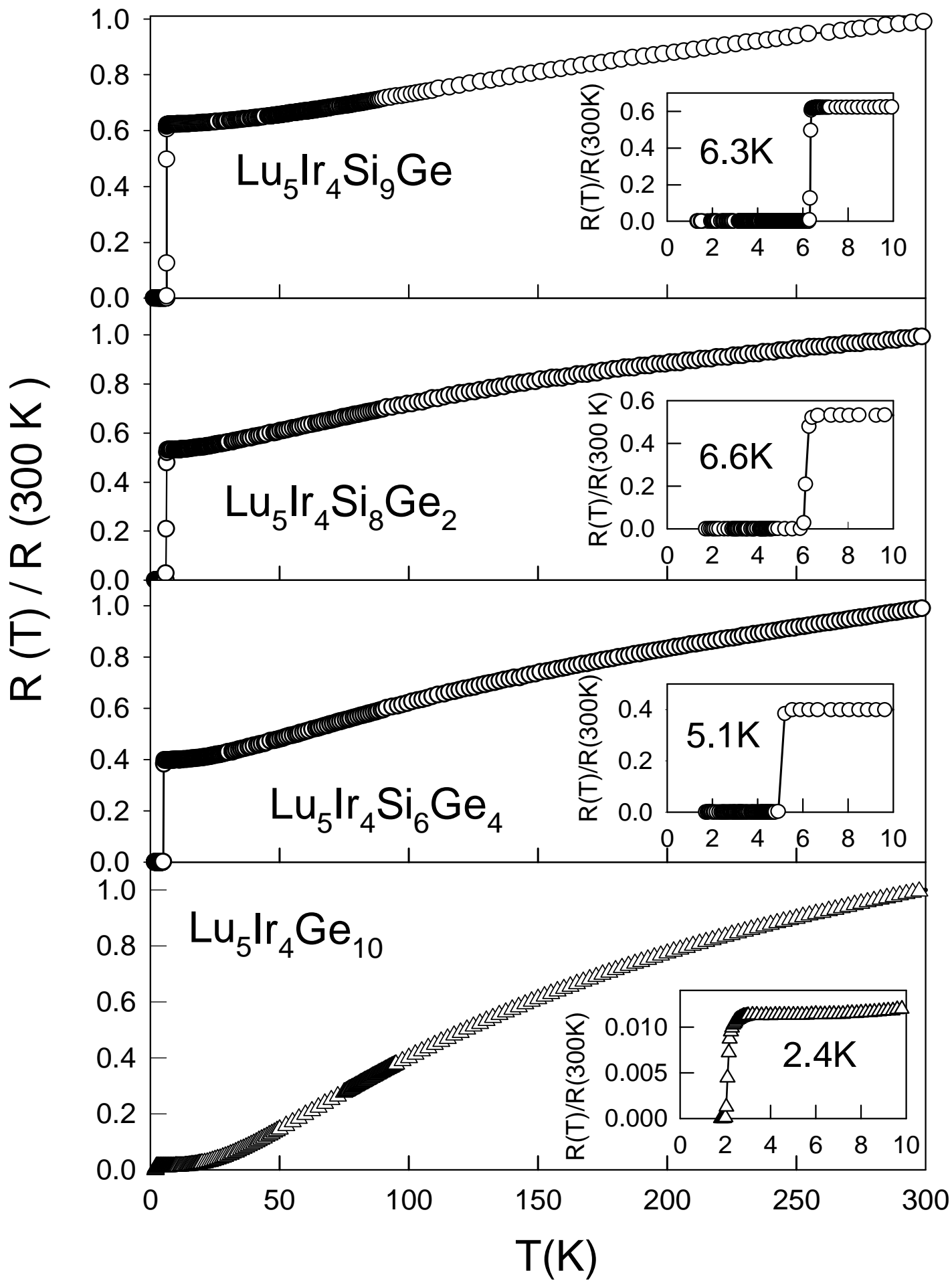
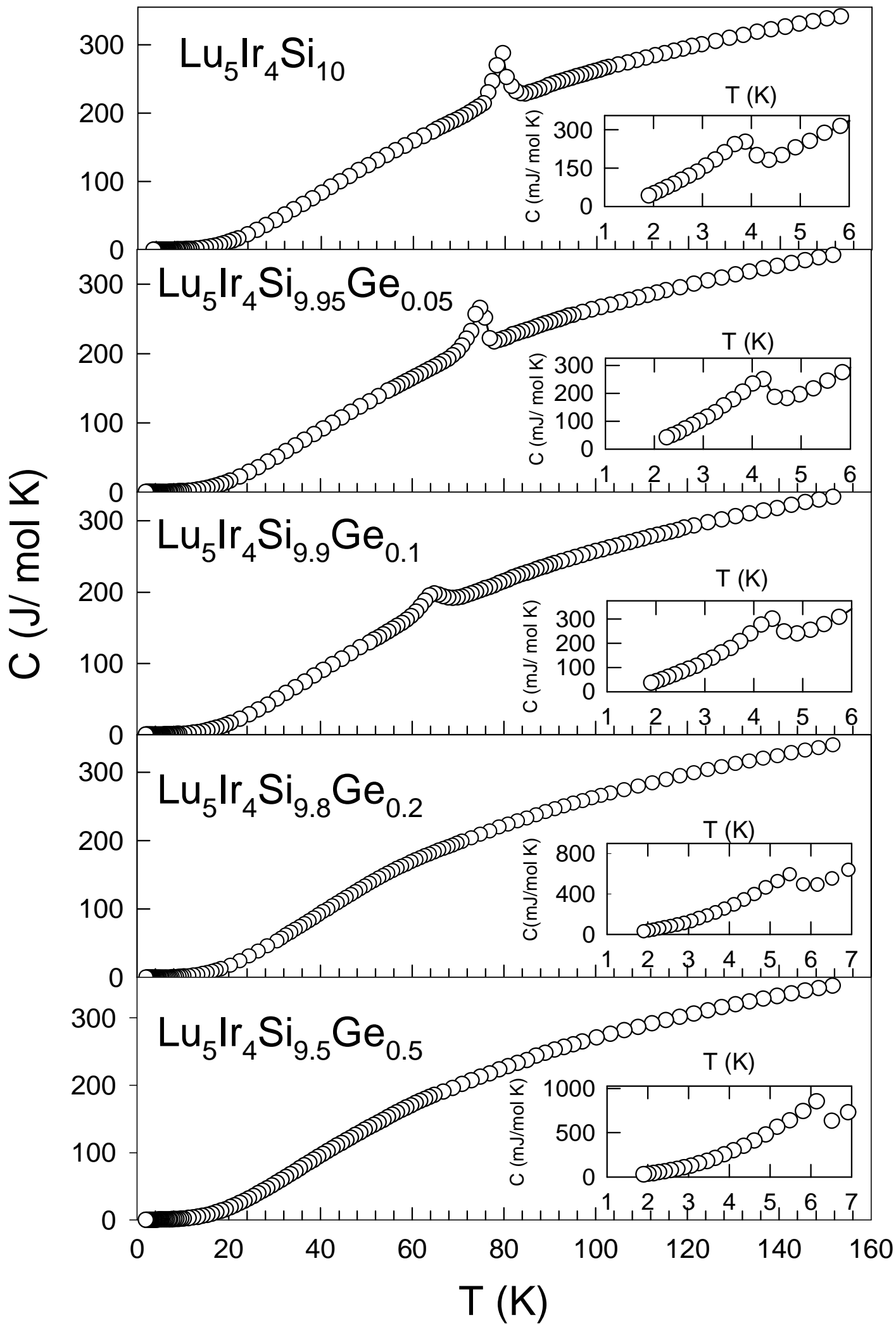
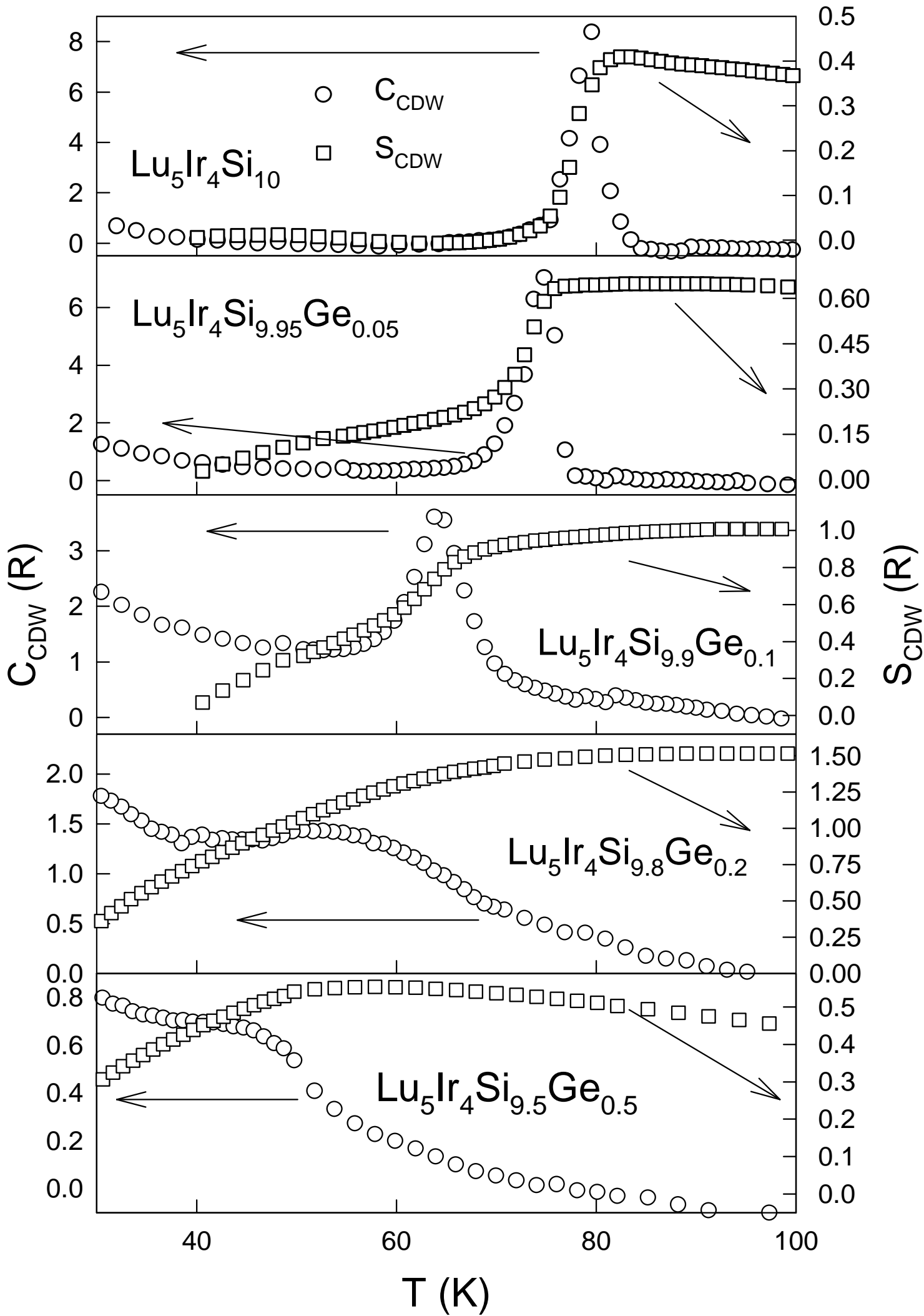
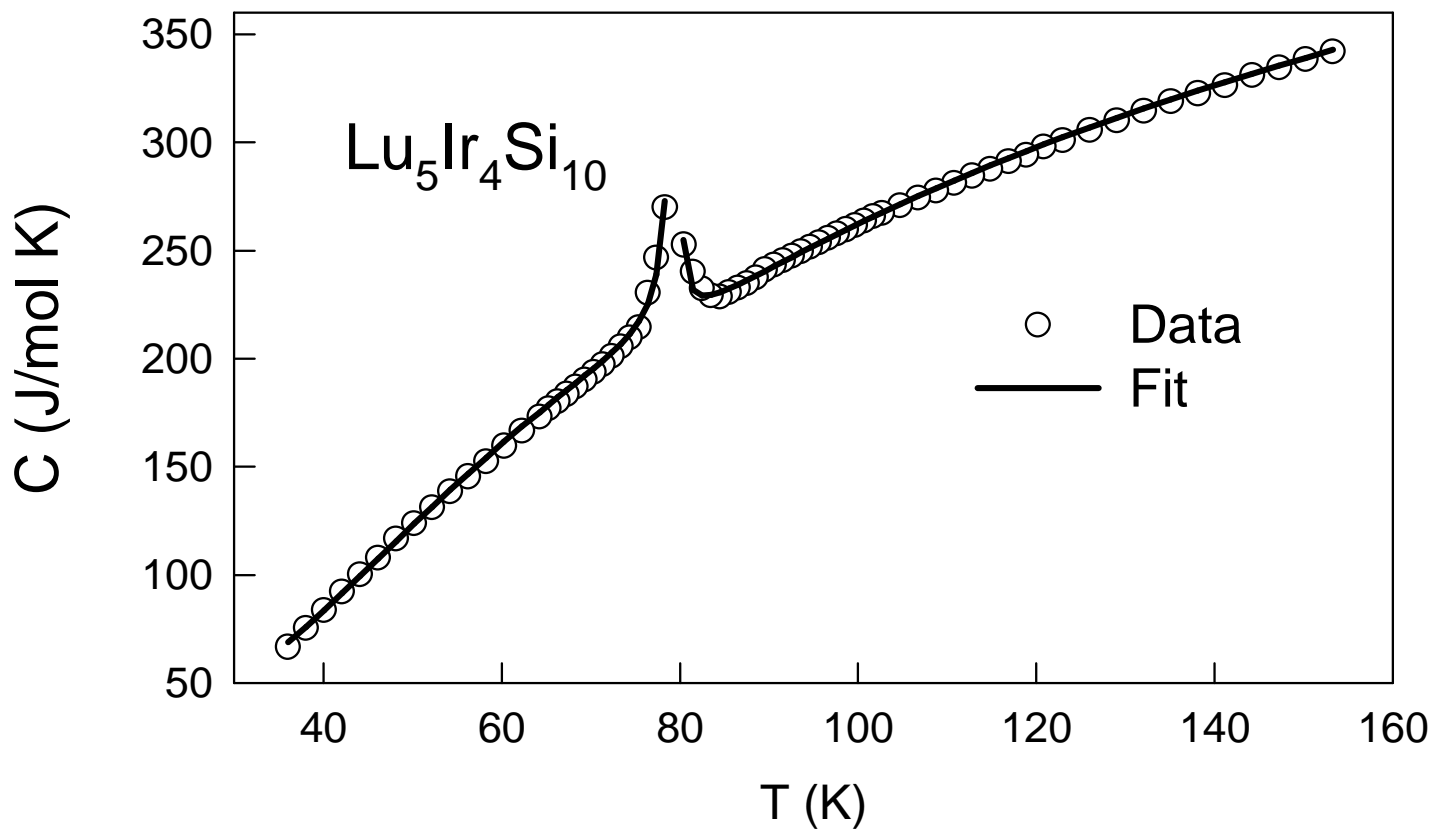
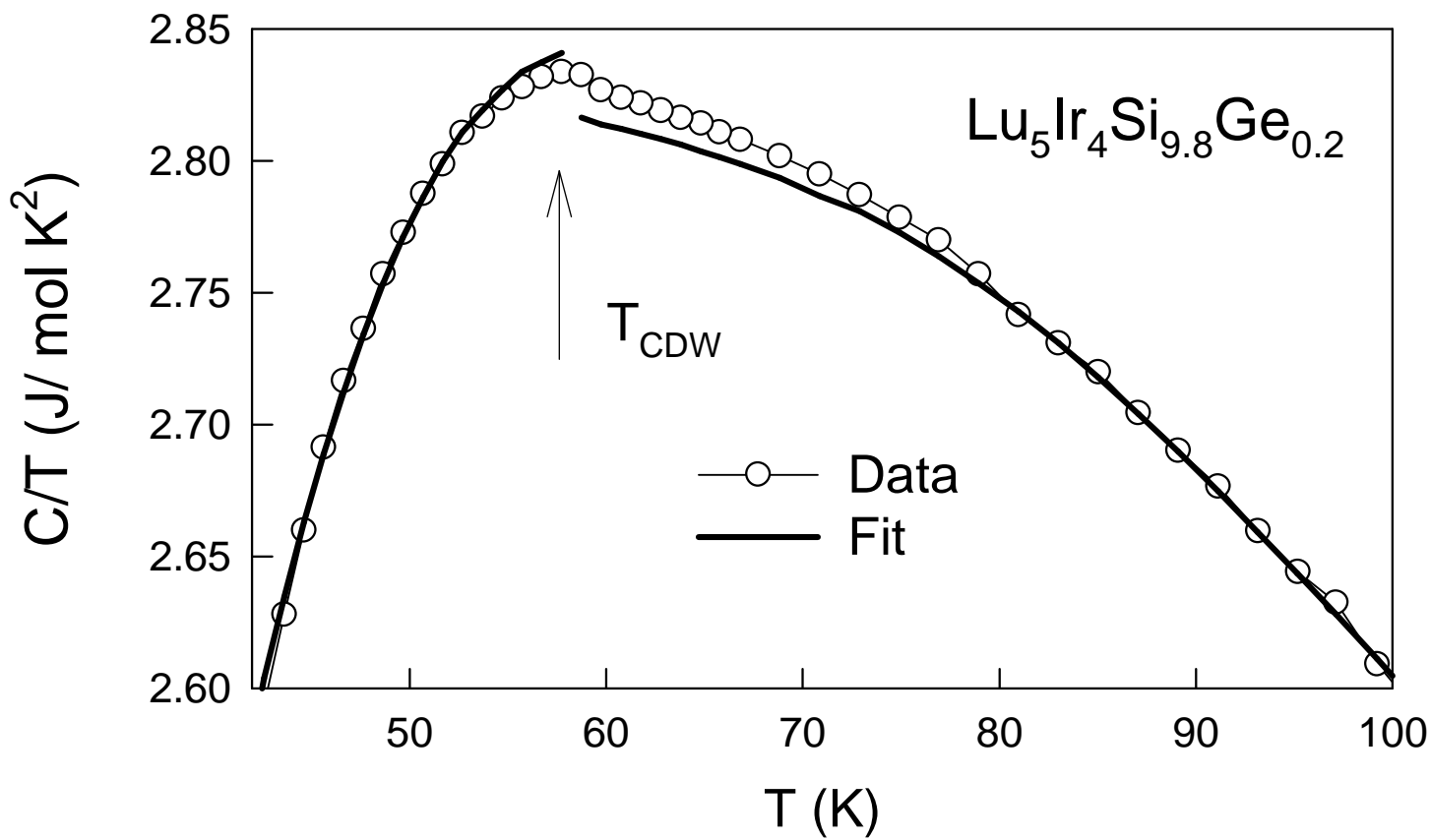
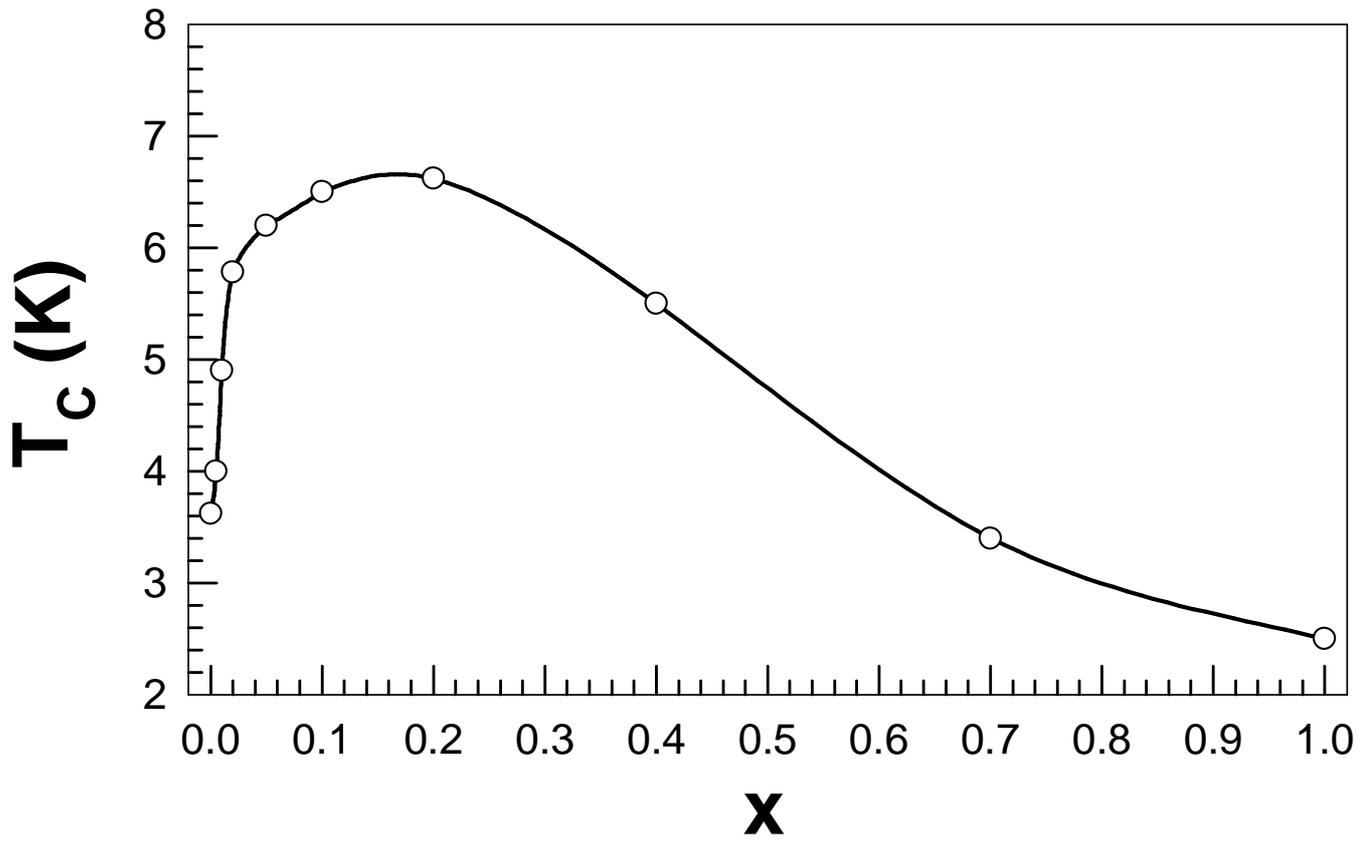
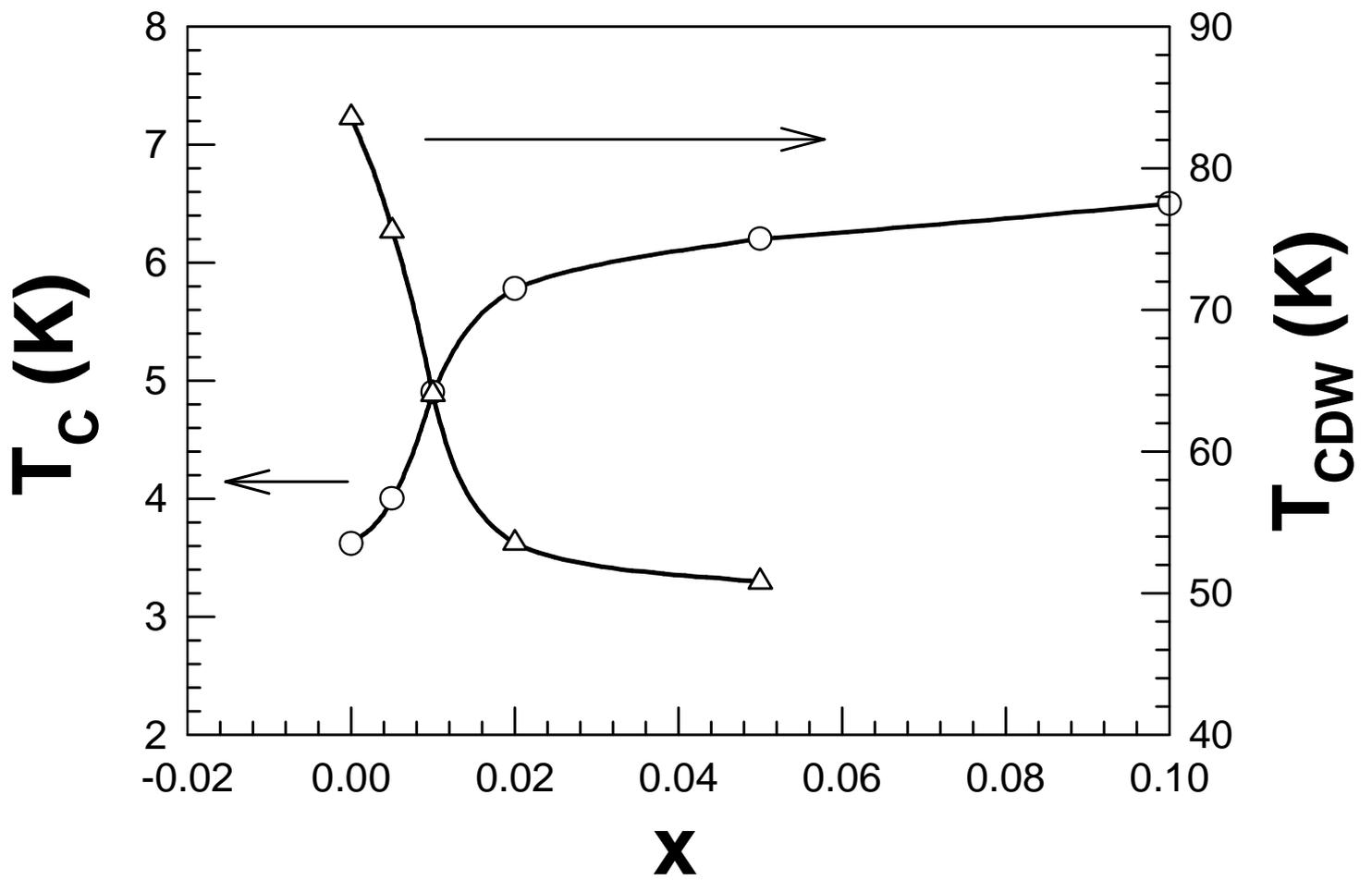
\end{document}